\documentclass[a4paper,11pt]{article}
\pdfoutput=1 

\usepackage{jcappub} 

\usepackage[T1]{fontenc}

\title{On Shapiro time delay in massive scalar-tensor theories}

\author[a]{P. I. Dyadina,}
\author[b]{ S. P. Labazova}

\affiliation[a]{Sternberg Astronomical Institute, Lomonosov Moscow State
University,\\ Universitetsky Prospekt, 13, Moscow 119991, Russia}
\affiliation[b]{Faculty of Mechanics and Mathematics, Lomonosov Moscow State University,\\ Leninskie Gory, 1, Moscow 119991, Russia}

\emailAdd{guldur.anwo@gmail.com}
\emailAdd{sp.labazova@physics.msu.ru}

\abstract{The problem of determining the post-Newtonian parameter $\gamma$ in massive scalar-tensor theories is considered. We demonstrate equivalent correspondence between the post-Newtonian parameter $\gamma$ and the parameter appearing in the equation of null geodysic in massive scalar-tensor theories. We show that massive scalar-tensor theories can be distinguished from general relativity based on the predictions of Shapiro time delay. All calculations are presented using hybrid metric-Palatini f(R)-gravity as an example. The expression for Shapiro time delay  in hybrid f(R)-gravity is obtained for the first time.}


\begin{document}
\maketitle
\flushbottom

\section{Introduction}
Initially, the parametrized post-Newtonian (PPN) formalism was invented as a way to unifiedly compare theories of gravity with each other and with experiment \cite{PPN1, PPN2, ppn1, will}. This formalism assumes the possibility of representing the metric of any metric theory of gravity in a certain form, in such a way that the metric will contain only post-Newtonian parameters and post-Newtonian potentials. The potentials do not change from theory to theory, while a set of 10 PPN parameters will be unique for each theory of gravity. Moreover, all PPN parameters can be obtained experimentally. Due to this fact, effective test of gravitational theories becomes possible. However, the PPN formalism was invented for theories that do not contain massive fields. In case of applying the PPN formalism to massive scalar-tensor theories of gravity, there is still ambiguity.

The point is that in massive theories of gravity, in addition to classical post-Newtonian potentials of the form $ 1/r $, Yukawa potentials are present \cite{alsing}. Thus, additional terms appear in the PPN metric that cannot be included in the PPN metric in its original form. It becomes necessary to modify the PPN parameters and introduce a dependence on the distance into them (while in classical PPN formalism PPN parameters are always constants), so that the PPN potentials retain their unchanged form  from theory to theory \cite{meppn}. And then the question arises, will these modified PPN parameters carry the same physical meaning as the previous ones? Will they be determined by the same experiments? In this article we will try to answer this question using the example of one single PPN  parameter $\gamma$.

The parameter $\gamma$  has a certain physical meaning in the classical PPN formalism: it is responsible for the effect of light deflection in the field of a massive object \cite{will, gamma}. In the case of massless theories this parameter appears in the equation of the null geodesic. The main question is whether the parameter standing in the equation of the null geodesic in massive theories of gravity coincides with the PPN parameter $\gamma$. Another question is whether such theories will predict the magnitude of the Shapiro delay different from general relativity.

This issue was particularly investigated in the context of massive scalar-tensor theories. The most well-known scalar-tensor theory is Brans-Dicke theory \cite{brans}. Now a massive version of this theory is of great interest, which is considered as one of the ways to describe the accelerated expansion of the Universe. L.~Perivolaropoulos in the paper \cite{papa} considered massive version of Brans-Dicke theory and applied PPN formalism to this model. He found that PPN parameter $\gamma$ depends not only on the model parameters of the theory but also on the radial distance from the Sun. On the other hand the Shapiro time delay was calculated in massive Brans-Dicke model in the work \cite{alsing} by J.~Alsing et al. The authors also found that the parameter in the null geodesic equation (which corresponds to parameter $\gamma$ for massless theories) will depend on the distance. In addition, in other work  \cite{deng} X.-M.~Deng and Y.~Xie also calculated Shapiro time delay in massive Brans-Dicke theory and showed that light deflection and Cassini tracking \cite{gamma} cannot distinguish massive scalar-tensor theory from general relativity (GR). In this work, authors are based on their result that $\gamma$ in massive scalar-tensor theories will be equal $\gamma_{GR}$ in GR ($\gamma_{GR}=1$). In all these works scalar-tensor theory with an arbitrary coupling function $\omega(\phi)$ and a generic potential $V(\phi)$  was considered.

In addition to the Brans-Dicke theory, in Horndeski gravity the light deflection was also calculated in two ways. In the work \cite{hou}, S.~Hou  and Y.~Gong found an explicit form of the PPN parameter $\gamma$, first from the post-Newtonian metric, and then from the expression for the Shapiro time delay. They obtained the same results in both cases, but since this was not the main goal of the article, so the authors did not pay much attention to the discussion of this result. Moreover, the PPN parameter $\gamma$ for Horndeski gravity was obtained earlier in the work \cite{hohman}.

In this article we will demonstrate equivalent correspondence between the PPN parameter $\gamma$ and the parameter appearing in the equation of null geodysic for one more massive scalar-tensor theory. For this purpose, we will consider hybrid metric-Palatini f(R)-gravity which can be represented as massive scalar-tensor theory \cite{hybrid1}. Moreover, we will show that massive scalar-tensor theories and general relativity predict different results regarding the magnitude of Shapiro delay in the gravitational field of a massive object.

	The hybrid metric-Palatini f(R)-gravity belongs to a family of f(R)-theories~\cite{fr1,fr2}. The action in f(R)-theories is constructed by generalizing  the gravitational part of the Einstein-Hilbert action as an arbitrary function of the curvature $R$. There are two possible approaches that can be used to obtain field equations from those modified actions: the metric one and the Palatini one.  In the metric approach  $g_{\mu\nu}$ is the only dynamical variable. Furthermore, the action is varied in respect to only $g_{\mu\nu}$. The Palatini method considers the idea of a  connection which defines the Riemann curvature tensor to be a priori independent of the metric.  Thus, variations with respect to the metric and the connection are performed independently. Additionally, the Palatini method provides second order differential field equations, while in the metric approach these equations are of the fourth order~\cite{capo1,capo2}. Unfortunately, both methods lead to some unsolvable  problems. The metric f(R)-theories, in general, cannot pass the standard Solar System tests~\cite{chiba,olmo1,olmo2}. All the Palatini f(R)-models aimed at explaining the accelerated Universe expansion lead to microscopic matter instabilities and to unacceptable features in the evolution patterns of cosmological perturbations~\cite{koivisto1,koivisto2}. Hybrid f(R)-gravity was constructed as a mixture of Palatini and metric f(R)-theories. It unites all of the advantages of both approaches but lacks their shortcomings \cite{hybrid2}.

In a recent work \cite{meppn} it was shown that the presence of the light scalar field in hybrid f(R)-gravity does not contradict the experimental data from Solar system. This is based not only on the $\gamma$ parameter, but also on all of the other parameters of the post-Newtonian formalism.  Thus, in contrast to metric f(R)-theory (excluding some metric f(R)-models \cite{odintsov1, odintsov2, odintsov3}) hybrid f(R)-gravity can pass the full post-Newtonian test.  Therefore, it will be interesting to show the equality of the $\gamma$ parameters obtained from the PPN metric directly and from the equation of the null geodesic for massive scalar-tensor theories.
	
The structure of the paper is the following. In section~\ref{sec2} we consider the action and the field equations of the hybrid metric-Palatini theory in a general form and in a scalar-tensor representation.  In section~\ref{sec3}, we calculate the PPN parameter $\gamma$ and Shapiro time delay in detail. We conclude in section~\ref{sec:conclusions} with a summary and discussion.
	
		Throughout this paper the Greek indices $(\mu, \nu,...)$ run over $0, 1, 2, 3$ and the signature is  $(-,+,+,+)$. We will be working in units $h = c = k_B = 1$ throughout the paper.
		
		\section{Hybrid f(R)-gravity}\label{sec2}
	
	The action of hybrid f(R)-gravity has the form~\cite{hybrid1,hybrid2}:
	\begin{equation}\label{act}
	S=\frac{1}{2k^2}\int d^4x\sqrt{-g}\left[R+f(\Re)\right]+S_m,
	\end{equation}
	where $k^2=8\pi G$, $G$ is the Newtonian gravitational constant, $R$ and $\Re=g^{\mu\nu}\Re_{\mu\nu}$ are the metric and Palatini curvatures respectively, $g$ is the metric determinant, $S_m$ is the matter action. Here the Palatini curvature $\Re$ is defined as a function of $g_{\mu\nu}$ and the independent connection $\hat\Gamma^\alpha_{\mu\nu}$:
	\begin{equation}\label{re}	
	\Re=g^{\mu\nu}\Re_{\mu\nu}	=g^{\mu\nu}\bigl(\hat\Gamma^\alpha_{\mu\nu,\alpha}-\hat\Gamma^\alpha_{\mu\alpha,\nu}+\hat\Gamma^\alpha_{\alpha\lambda}\hat\Gamma^\lambda_{\mu\nu}-\hat\Gamma^\alpha_{\mu\lambda}\hat\Gamma^\lambda_{\alpha\nu}\bigr).
	\end{equation}
	
	Like in the pure metric and Palatini cases, the hybrid f(R)-gravity~(\ref{act}) can be rewritten in a scalar-tensor representation (for details see~\cite{hybrid1,hybrid2}):
	\begin{equation}\label{stact1}
	S=\frac{1}{2k^2}\int d^4x\sqrt{-g}\biggl[(1 + \phi)R + \frac{3}{2\phi}\partial_\mu \phi \partial^\mu \phi - V(\phi)\biggr]+S_m,	
	\end{equation}
	where $\phi$ is a scalar field and $V(\phi)$ is a scalar field potential. Here and further we use the Jordan frame.
	
	Then the metric and scalar field equations take the following forms~\cite{hybrid1,hybrid2}:
	\begin{align}	
	\label{feh}
	&(1+\phi)R_{\mu\nu}=k^2\left(T_{\mu\nu}-\frac{1}{2}g_{\mu\nu}T\right)+\nabla_\mu\nabla_\nu\phi+\frac{1}{2}g_{\mu\nu}\biggl[V(\phi)+\nabla_\alpha\nabla^\alpha\phi\biggr]-\frac{3}{2\phi}\partial_\mu\phi\partial_\nu\phi,\\ 	
	\label{fephi}
		&\nabla_\mu\nabla^\mu\phi-\frac{1}{2\phi}\partial_\mu\phi\partial^\mu\phi-\frac{\phi[2V(\phi)-(1+\phi)V_\phi]}{3}=-\frac{k^2}{3}\phi T, 
\end{align}
	where $T_{\mu\nu}$ and $T$ are the energy-momentum tensor and its trace respectively.
	
	\section{Light deflection}\label{sec3}
	\subsection{PPN parameter $\gamma$}
	Firstly we show the calculation  of PPN parameter $\gamma$ in hybrid f(R)-gravity from PPN metric directly. To achieve this goal, it is necessary to solve the field equations in the PPN approximation up to the order $O(2)$ \cite{will,will1,will2}.
To obtain the linearized field equations in the weak-field limit we consider the following perturbations of the scalar field and
metric tensor:
	\begin{equation}\label{decompos}
	\phi=\phi_0+\varphi,\qquad\ g_{\mu\nu}=\eta_{\mu\nu}+h_{\mu\nu},
	\end{equation}
	where $\phi_0$ is the asymptotic background value of the scalar field far away from the source, $\eta_{\mu\nu}$ is the Minkowski background, $h_{\mu\nu}$ and $\varphi$ are the small perturbations of tensor and scalar fields respectively. Here we consider $\phi_0$ as a constant. The scalar potential $V(\phi)$ could be expanded in a Taylor series around the background value of the scalar field $\phi_0$ like
	\begin{equation}\label{V}
	V(\phi)=V_0+V'\varphi+\frac{V''\varphi^2}{2!}+\frac{V'''\varphi^3}{3!}...,
	\end{equation}
	hence its derivative with respect to $\varphi$ will be the following: $V_\phi=V'+V''\varphi+V'''\varphi^2/2$.
	
	The field equation for the scalar field~(\ref{fephi}) in the leading perturbation order ($O(2)$) takes the form~\cite{hybrid1,hybrid2,meppn}:
	\begin{equation}\label{fephi2}
	\left(\nabla^2-m_\varphi^2\right)\varphi^{(2)}=-\frac{k^2\phi_0}{3}T,
	\end{equation}
		where we denote $m_\varphi^2=[2V_0-V'-(1+\phi_0)\phi_0V'']/3$ as a scalar field mass.	The superscript $^{(2)}$ indicates the order of perturbation.
		
	The linearized equations for the metric is given by~\cite{hybrid1,hybrid2}
	\begin{equation}\label{feh002}
	-\frac{1}{2}\nabla^2h_{\mu\nu}=-\frac{k^2}{(1+\phi_0)}\biggl(T_{\mu\nu}-\frac{1}{2}T\eta_{\mu\nu}\biggr)+\frac{V_0+\nabla^2\varphi}{2(1+\phi_0)}\eta_{\mu\nu},
	\end{equation}
 To obtain this result we used  Nutku gauge conditions~\cite{Nutku}:
	\begin{equation}\label{gauge}
	h^\alpha_{\beta,\alpha}-\frac{1}{2}\delta^\alpha_\beta h^\mu_{\mu,\alpha}=\frac{\varphi_{,\beta}}{1+\phi_0}.
	\end{equation}

Thus, the leading order of metric perturbations is defined as~\cite{hybrid2,meppn}:	
	\begin{align}
	\label{h002_1}
	h_{00}^{(2)}=\frac{k^2}{4\pi(1+\phi_0)}\frac{M}{r}\left(1-\frac{\phi_0}{3}  \rm e^{-m_\varphi r}\right)+\frac{V_0}{1+\phi_0}\frac{r^2}{6},\\
	\label{hij_1}
	h_{ij}^{(2)}=\frac{\delta_{ij}k^2}{4\pi(1+\phi_0)}\frac{M}{r}\left(1+\frac{\phi_0}{3}  \rm e^{-m_\varphi r}\right)-\delta_{ij}\frac{V_0}{1+\phi_0}\frac{r^2}{6},
	\end{align}
where $M$ is the Solar mass and $\delta_{ij}$ is the Kronecker delta. Here $V_0/(\phi_0+1)$ is the cosmological constant term. It must be negligible in Solar System scales in order not to affect the local dynamics \cite{blackhole}. That is why further this term is not considered \cite{meppn}. 

Since $h_{00}$  up to the order $O(2)$ in the general post-Newtonian metric is defined as \cite{will,will1,will2}
\begin{equation}\label{h00_2}
h^{(2)}_{00}=2\frac{G^{\rm eff}M}{r},
	\end{equation}
 it is possible to extract the effective gravitational constant from~(\ref{h002_1})~\cite{hybrid1,hybrid2,meppn}:
	\begin{equation}\label{Geff}
	G^{\rm eff}=\frac{k^2}{8\pi(1+\phi_0)}\left(1-\frac{\phi_0}{3} \rm e^{-m_\varphi r}\right).
	\end{equation}
Throughout the article we add the superscript $^{\rm eff}$ to the PPN parameters which are considered as spatially dependent functions. 

After that we can extract PPN parameter $\gamma^{\rm eff}$ from the expression~(\ref{hij_1}). The metric pertubation $h_{ij}$ up to the order $O(2)$ in the general post-Newtonian metric has the following form\cite{will,will1,will2}:
\begin{equation}\label{hij_2}
h^{(2)}_{ij}=2\gamma^{\rm eff}\frac{G^{\rm eff}M}{r}\delta_{ij}.
\end{equation}
	Therefore, from the equation~(\ref{hij_1}) we can express $\gamma^{\rm eff}$~\cite{hybrid1,hybrid2,meppn}:
	\begin{equation}\label{gamma}
	\gamma^{\rm eff}=\frac{1+\phi_0 \rm e^{-m_\varphi r}/3}{1-\phi_0 \rm e^{-m_\varphi r}/3}.
	\end{equation}
	As we can see, in hybrid f(R)-gravity, the PPN  parameter $\gamma^{\rm eff}$ depends on the distance as in the massive Brans-Dicke theory \cite{alsing}.  This was expected result since both theories are massive scalar-tensor models. In addition, the question of applying the PPN formalism to massive scalar-tensor theories and the dependence of PPN parameters on distance was considered in the recent paper \cite{meppn}.

	\subsection{Shapiro time delay}
	Secondly, we calculate the Shapiro time delay in the case of hybrid metric-Palatini f(R)-gravity.  The equation of a photon motion along a null geodesic has the form:
	\begin{equation}\label{geodesic}
		g_{\mu\nu}u^\mu u^\nu=0,
	\end{equation}
	where $u^{\mu}=d x^{\mu}_a/d \tau_a$ is four-velocity of a photon, $\tau_a$ is the proper time of particle $a$ measured
along its worldline $ x^{\mu}_a$
	
	This equation can be expressed up to the order $O(2)$ as:
	\begin{equation}\label{geo}
		-1+h_{00}^{(2)}+\left(\delta_{ij}+h_{ij}^{(2)}\right)u^iu^j=0.
	\end{equation}
	Substituting~(\ref{h002_1}) and ~(\ref{hij_1})  into \eqref{geo}, the equation~(\ref{geodesic}) takes the following form:
	\begin{equation}\label{u3}
		-1+\frac{k^2}{4\pi(1+\phi_0)}\left(1-\frac{\phi_0}{3} \rm e^{-m_\varphi r}\right)\frac{M}{r}+\Bigg(1+\frac{k^2}{4\pi(1+\phi_0)}\left(1+\frac{\phi_0}{3}  \rm e^{-m_\varphi r}\right)\frac{M}{r}\Bigg)|\vec{u}|^2=0.
	\end{equation}
	If the photon was emitted at the point $\mathbf{x_e}$ in the direction of $\mathbf{n}$ at the time $t_e$, then its trajectory  is described by the expression:
		\begin{equation}\label{xi(t)}
		x^i(t)=x_e^i+n^i\left(t-t_e\right)+x^i_{PN}(t),
	\end{equation}
	taking into account the post-Newtonian corrections $x^i_{PN}(t)$.
Substituting this expression into \eqref{u3}, we obtain the following equation:
	\begin{equation}\label{u2}
		\mathbf{n}\cdot\frac{d\mathbf{x}_{PN}(t)}{dt}=\frac{dx_{PN}^{\parallel}(t)}{dt}=-\frac{k^2}{8\pi(1+\phi_0)}\frac{M}{r}.
	\end{equation}
	
Then the time of  photon traveling from the $\mathbf{x_e}$ to $\mathbf{x}$ and back will be equal to
	\begin{equation}
		\Delta t=2|\mathbf{x}-\mathbf{x_e}|-\frac{k^2}{8\pi(1+\phi_0)}\int\limits_{t_e}^t\frac{M}{r(t')}dt'.
	\end{equation}	

The second term on the right-hand side of this equation is the correction $\delta t$ due to the Shapiro delay.
It can be obtained after integrating:
	\begin{equation}\label{deltat}
		\delta t=4\frac{k^2M}{8\pi (1+\phi_0)}\ln{\Bigg[\frac{(r_e+\vec{r_e}\cdot\vec{n})(r_p-\vec{r_p}\cdot\vec{n})}{r_b^2}\Bigg]},
	\end{equation}
where the photon is emitted from $\bf{r}_e$ in the direction of $\bf{n}$, traveling to $\bf{r}_p$ and back again, $M$ is the mass of the body causing the time-delay and $\bf{r}_b$ is the impact parameter of the photon with respect to the source. The most important point is the mass appearing in \eqref{deltat} is not a measurable quantity \cite{alsing}. The actually
measured quantity is the Keplerian mass:
	\begin{equation}
		M_k=G_{\rm eff}M=M\frac{k^2}{8\pi(1+\phi_0)}\left(1-\frac{\phi_0}{3} \rm e^{-m_\varphi r}\right),
	\end{equation}
where $r$ should be thought of as a fixed quantity which depends on how the Keplerian mass of the body was determined. For example, in the case of the Solar System $r$ should be set to 1 AU, since this is the scale associated with the determination of the Keplerian mass of the Sun. In terms of $M_k$ we find
	\begin{equation}
		\delta t=\frac{4M_k}{1-\frac{\phi_0}{3} \rm e^{-m_\varphi r}}\ln{\Bigg[\frac{(r_e+\vec{r_e}\cdot\vec{n})(r_p-\vec{r_p}\cdot\vec{n})}{r_b^2}\Bigg]}.
	\end{equation}
	
	This equation can be expressed as
	\begin{equation}
		\delta t=2M_k\left(1+\tilde{\gamma}\right)\ln{\Bigg[\frac{(r_e+\vec{r_e}\cdot\vec{n})(r_p-\vec{r_p}\cdot\vec{n})}{r_b^2}\Bigg]},
	\end{equation}
	where
		\begin{equation}\label{gammanew}
	\tilde{\gamma}=\frac{1+\phi_0 \rm e^{-m_\varphi r}/3}{1-\phi_0 \rm e^{-m_\varphi r}/3}.
	\end{equation}
	Thus we obtain that
	\begin{equation}
	\tilde{\gamma}=\gamma^{\rm eff}
	\end{equation}
	in hybrid f(R)-gravity. It is worth noting that ``r'' is the distance at which the Keplerian mass of the Sun is measuring in both parameters $\tilde{\gamma}$ and $\gamma^{\rm eff}$. Therefore, the deflection of a light ray by the gravitational field of the Sun depends not only on the mass of the object but also on the model parameters and the distance $r$. This is an important difference between massive scalar-tensor theories of gravity and general relativity. It should be said separately that the expression for Shapiro time delay  in hybrid f(R)-gravity was obtained in this work for the first time. The restrictions imposed by $\gamma$ parameter on the parameters of the theory were obtained earlier in the works \cite{meppn, leaniz}.
	
	It is important to emphasize that the expression of the Shapiro delay \ref{deltat} contains the mass of an gravitating object. For calculating the value of Shapiro delay, it is necessary to use the experimental values of the quantities included in the theoretical expression. The mass of an gravitating object can be obtained from Kepler's third law, which is modified in massive theories of gravity \cite{pulsars}. Instead of the Newtonian gravitational constant Kepler's third law contains an effective gravitational constant. We cannot use the ``true'' mass of the object, since it is unknown from the experiment. Then it is necessary to change the mass in the expression \ref{deltat} for the mass from the experiment and  in accordance with Kepler's third law it is $G_{eff}M$. It is the key difference between the our work and the article \cite{deng}, since the authors in their paper \cite{deng} did not take into account the necesserity to use the Keplerian mass for calculating Shapiro delay.
	
	\section{Conclusions and discussion}\label{sec:conclusions}
	In this investigation, we posed two questions:  whether the parameter standing in the equation of the null geodesic in massive scalar-tensor theories of gravity coincides with the PPN parameter $\gamma$ and whether such theories will predict the magnitude of the Shapiro delay different from general relativity. We discussed these issues using the hybrid metric-Palatini f(R)-gravity which can be represented as massive scalar-tensor theory \cite{hybrid1,hybrid2}. In the framework of the considered theory, we showed that  the post-Newtonian parameter $\gamma^{\rm eff}$ calculated from the post-Newtonian metric directly is equal to  the parameter $\tilde{\gamma}$ obtained from the expression for the Shapiro time delay of a light ray passing near a massive body. Therefore, in the case of hybrid f(R)-gravity  the parameters are identical. In addition, in this work, for the first time, Shapiro time delay  in hybrid f(R)-gravity was obtained in this work for the first time.
	
	Previously, similar studies were conducted in the framework of the massive Brans-Dicke theory. In the paper \cite{papa} the post-Newtonian parameter $\gamma$  was calculated from the post-Newtonian metric and later in \cite{alsing} the Shapiro time delay was investigated. Similar to the case of hybrid f(R)-theory, both methods had led to the same result. Later on, this problem was considered in \cite{deng}. Authors had also calculated the Shapiro time delay in the massive Brans-Dicke theory. However, they found that the light deflection and Cassini tracking \cite{gamma} cannot distinguish a massive scalar-tensor theory from GR. This conclusion is based on the result obtained by the authors that the $\gamma$ parameter in massive Brans-Dicke theory is equal to the $\gamma_{GR}=1$. The authors also demonstrate that the Shapiro delay is determined only by the mass of the object and does not depend on the parameters of the massive theory. These conclusions are not consistent with result obtained earlier in the paper \cite{alsing} and in our work. The difference in results is that in the latter work \cite{deng} it was not taken into account an inequality of the observed Keplerian mass and mass of the body causing the time-delay. 
	
	In the paper \cite{hou} authors had obtained  expressions for PPN parameter $\gamma$ using the two methods that were previously discussed in Horndeski theory. Both expressions were identical (see eq. (38) and eq.(50)-(51) in the paper \cite{hou}). Since  Horndeski gravity is the most general scalar-tensor theory providing the second-order field equations which evades Ostrogradski instabilities, this equality can be applied for all special cases of the Horndeski theory. This fact was demonstrated on the examples of Brans-Dicke theory and hybrid f(R)-gravity.
	
	Based on the foregoing, we can conclude that both massless and massive scalar-tensor theories can be distinguished from GR based on the data of  light deflection by the gravitational field of the Sun. In addition when considering the  massive scalar-tensor theory light deflection will be determined by the PPN parameter $\gamma$ as in the case of the massless theory. However, the difference between these models is that the parameter $\gamma$  in massive scalar-tensor theories depends on the distance at which the Keplerian mass was measured. Therefore, it can be argued that the PPN formalism gives the same predictions as the direct calculation of the Shapiro time delay and is applicable to the massive scalar-tensor theories.
	
	This work is the first step in obtaining the universal apparatus for testing gravitational theories with massive fields in the weak-field limit as the original PPN formalism for massless, but this is a topic of more extensive research in the future.

\acknowledgments

The authors thank N. A. Avdeev for discussions and comments on the topics of this paper. The work was supported by the Foundation for the Advancement of Theoretical Physics and Mathematics “BASIS”.

\end{document}